\documentclass{ieeeaccess}
\usepackage{cite}
\usepackage{amsmath,amssymb,amsfonts}
\usepackage{algorithmic}
\usepackage{graphicx}
\usepackage{textcomp}
\usepackage{booktabs} 
\usepackage{threeparttable}
\usepackage{booktabs}
\usepackage{bm}
\makeatletter
\AtBeginDocument{\DeclareMathVersion{bold}
\SetSymbolFont{operators}{bold}{T1}{times}{b}{n}
\SetSymbolFont{NewLetters}{bold}{T1}{times}{b}{it}
\SetMathAlphabet{\mathrm}{bold}{T1}{times}{b}{n}
\SetMathAlphabet{\mathit}{bold}{T1}{times}{b}{it}
\SetMathAlphabet{\mathbf}{bold}{T1}{times}{b}{n}
\SetMathAlphabet{\mathtt}{bold}{OT1}{pcr}{b}{n}
\SetSymbolFont{symbols}{bold}{OMS}{cmsy}{b}{n}
\renewcommand\boldmath{\@nomath\boldmath\mathversion{bold}}}
\makeatother

\usepackage{multirow}
\usepackage[table,xcdraw]{xcolor}
\usepackage[normalem]{ulem}
\useunder{\uline}{\ul}{}
\usepackage{adjustbox}

\def\BibTeX{{\rm B\kern-.05em{\sc i\kern-.025em b}\kern-.08em
    T\kern-.1667em\lower.7ex\hbox{E}\kern-.125emX}}

\begin{document}
\history
\doi

\title{Empowering Few-Shot Recommender Systems with Large Language Models-Enhanced Representations}
\author{\uppercase{Zhoumeng Wang}\authorrefmark{1}}
\address[1]{The Chinese University of Hong Kong (e-mail: johnnywang@link.cuhk.edu.hk)}
\tfootnote{}

\markboth
{Wang: Empowering Few-Shot Recommender Systems with Large Language Models-Enhanced Representations}
{Wang: Empowering Few-Shot Recommender Systems with Large Language Models-Enhanced Representations}

\corresp{Corresponding author: Zhoumeng Wang (e-mail: johnnywang@link.cuhk.edu.hk).}

\begin{abstract}
Recommender systems utilizing explicit feedback have witnessed significant advancements and widespread applications over the past years. However, generating recommendations in few-shot scenarios remains a persistent challenge. Recently, large language models (LLMs) have emerged as a promising solution for addressing natural language processing (NLP) tasks, thereby offering novel insights into tackling the few-shot scenarios encountered by explicit feedback-based recommender systems. To bridge recommender systems and LLMs, we devise a prompting template that generates user and item representations based on explicit feedback. Subsequently, we integrate these LLM-processed representations into various recommendation models to evaluate their significance across diverse recommendation tasks. Our ablation experiments and case study analysis collectively demonstrate the effectiveness of LLMs in processing explicit feedback, highlighting that LLMs equipped with generative and logical reasoning capabilities can effectively serve as a component of recommender systems to enhance their performance in few-shot scenarios. Furthermore, the broad adaptability of LLMs augments the generalization potential of recommender models, despite certain inherent constraints. We anticipate that our study can inspire researchers to delve deeper into the multifaceted dimensions of LLMs’ involvement in recommender systems and contribute to the advancement of the explicit feedback-based recommender systems field.
\end{abstract}


\titlepgskip=-21pt

\maketitle

\section{Introduction}
\label{sec:introduction}
\PARstart{R}{ecommender} systems are defined as techniques that utilize users’ explicitly or implicitly expressed preferences to provide recommendations for items of interest, address the issue of information overload, and deliver novelty and surprise\cite{01mocean2012marketing}. With the advancement of deep learning, the field of recommender systems has witnessed significant progress in recent years. Initially, collaborative filtering and ID-based methods are widely adopted across diverse recommendation scenarios\cite{02bobadilla2020deep}\cite{03rezaimehr2021survey}\cite{04zhang2019deep}. Subsequently, there has been a growing research focus on incorporating textual side information into recommender systems to develop knowledge-based\cite{05cena2021logical}\cite{06dong2020interactive}\cite{07alamdari2020systematic} and content-based\cite{08mittal2020smart}\cite{09perez2021content} approaches that effectively leverage explicit feedback.

However, the majority of recommendation methods continue to grapple with multiple long-standing challenges. The mobile nature of cyber users and the continuous emergence of new items have underscored the significance of few-shot scenarios, where recommender systems are required to provide recommendations based on limited user information. Simultaneously, recommender systems commonly possess a task-specific property that constrains their generalization capabilities across different data sources and application scenarios. Such property is currently being challenged in the dynamic cyberspace, where explicit feedback from users has become increasingly complex and overwhelming in volume. Moreover, as essential tools for consumer engagement, marketing, and business analysis\cite{10ansari2000internet}\cite{11bodapati2008recommendation}, recommender systems necessitate interpretability and transparency; nevertheless, the integration of deep learning has hindered these aspects.

The recent advancements in large language models (LLMs) have offered promising prospects for addressing the aforementioned challenges. Emerging LLMs with generative and logical reasoning capabilities, such as ChatGPT, exhibit remarkable proficiency in text summarization and possess potential for association\cite{12brown2020language}\cite{13dai2023chataug}, thereby endowing them with a natural aptitude for engagement in textual explicit feedback processing. Meanwhile, the integration of LLMs into diverse recommendation tasks from various perspectives has emerged as a pivotal area of investigation. Nevertheless, prior research\cite{14liu2023chatgpt} suggests that when employed directly and solely as a recommender system in few-shot scenarios, LLMs do not demonstrate superior performance across various tasks compared to traditional recommendation models. In contrast, recent studies highlight LLMs' effective participation in recommendations as a component of recommender systems\cite{15di2023evaluating}\cite{16gao2023chat}. This motivates our novel research proposal: investigating the potential of utilizing LLMs to generate user and item representations using textual explicit feedback, thereby enhancing the performance of existing recommender models in few-shot scenarios. 

To investigate this subject, we conduct an in-depth study by referencing previous research\cite{14liu2023chatgpt}\cite{17kefato2021dynamic}. We develop a template to process movie reviews from a deliberately selected public dataset using LLMs to generate user and item representations. These representations are then incorporated into selected recommendation models for evaluation on two tasks: interaction prediction and direct recommendation. To specifically investigate the extraction and association capabilities of the experimental LLMs, we manually adjusted the number of training samples to simulate a few-shot scenario. 

Comprehensive experimental results indicate that utilizing LLMs for representation generation significantly enhances the performance of specific recommendation models in a few-shot scenario, demonstrating that LLMs can effectively serve as an explicit feedback processing method for multiple recommendation tasks. Our manual observations also suggest that certain LLMs with generative and logical reasoning capabilities possess a distinctive ability to generate supplementary information through association. LLMs’ broad applicability across diverse scenarios and proficiency in processing textual information even in the absence of quantitative metrics can augment the generalization potential of recommender systems. It is worth noting that the observed enhancements are more pronounced in recommendation models that integrate neural networks. This phenomenon could be attributed to inherent constraints imposed by model structures and characteristics of the embeddings. 

We hope the results of this experiment can inspire researchers to further explore the incorporation of LLMs into the recommendation process, while offering valuable insights in specific research fields, such as interpretability, cold-start challenges, and model enhancement within explicit feedback-involved recommender systems.

\section{RELATED STUDY}
\subsection{Explicit Feedback for Recommendation}
In contrast to implicit feedback derived primarily from user behavior observations, explicit feedback is openly and actively provided by users themselves to reflect their preferences and attitudes. The concept of explicit feedback mentioned in the book \emph{Recommender Systems: An Introduction} encompasses ratings and annotations\cite{18jannach2010recommender}, while Konstan and Riedl\cite{19konstan2012recommender} broaden its definition to include diverse forms of user-contributed content such as reviews, tags, blog posts, tweets, Facebook updates, among others.

In previous studies, ratings have been regarded as a crucial form of explicit feedback that enhances the performance of recommender systems\cite{20zhao2018explicit}\cite{21liu2022ipr} and can be combined with implicit feedback to cater to diverse recommendation tasks\cite{22liu2010unifying}\cite{23jawaheer2010comparison}. Text, serving as another manifestation of explicit feedback, can also be leveraged by recommender systems. Textual explicit feedback is commonly manifested as user reviews and comments\cite{24betancourt2020use} that are generated in various languages\cite{25miao2017recommendation}. Other forms of textual explicit feedback include but are not limited to Tweets\cite{26chen2012collaborative}, web chats\cite{27loh2003tourism}, messages accompanied by geographic information\cite{28rosa2018knowledge}, and Tags\cite{29krestel2009latent}. Therefore, natural language processing (NLP) plays a crucial role in constructing recommender systems that rely on textual explicit feedback. Text mining has long been considered as an essential prerequisite in various recommendation models\cite{24betancourt2020use}, encompassing techniques such as Latent Dirichlet Allocation (LDA)\cite{30jakob2009beyond}, TF-IDF\cite{29krestel2009latent}, word segmentation\cite{25miao2017recommendation}, rule-based classifiers\cite{31li2010contextual}, and more. The processed text can be leveraged to support recommender systems built through approaches such as collaborative filtering\cite{25miao2017recommendation} \cite{30jakob2009beyond}, content-based filtering\cite{27loh2003tourism} or knowledge-based\cite{28rosa2018knowledge}. 

In recent years, the embedding process has emerged as a prominent focus in recommendation studies due to advancements in related research. The utilization of LLMs in recommendation has been increasingly prevalent owing to their proficiency in comprehending and processing human natural language\cite{32fan2023recommender}. Transformer architecture models (\emph{e.g.}, BERT, GPT, and T5\cite{33vaswani2017attention}) have been extensively employed in aspects including Pre-training, Fine-tuning, and Prompting \cite{32fan2023recommender}. Attention mechanism has also been integrated in the development of recommender system models. For instance, NARRE\cite{34chen2018neural}, a neural attention recommendation framework utilizing user reviews, is introduced to simultaneously predict users’ ratings towards items and generate review-level explanations for the prediction. Other attention models such as TARMF\cite{35lu2018coevolutionary} and MPCN\cite{36tay2018multi} that leverage textual explicit feedback also exhibit superior performance across diverse recommendation tasks compared to existing deep learning-based recommendation models (\emph{e.g.}, ConvMF\cite{37kim2016convolutional}, DeepCoNN\cite{38zheng2017joint}).

\subsection{ChatGPT for Recommendation}
Released by OpenAI in 2022, ChatGPT\cite{39openai2023gpt} is an advanced LLM and dialogue system that has demonstrated exceptional performance across various vertical domains. It showcases remarkable capabilities in context-based comprehension, summarization, and text generation\cite{12brown2020language}. The investigation into the methodology of transferring and employing ChatGPT’s extensive knowledge and paradigm acquired from large-scale corpora to recommendation scenarios has emerged as a cutting-edge pursuit in the academic domain.

ChatGPT can independently serve as a versatile recommendation model capable of handling various recommendation tasks. Liu \emph{et al.}\cite{14liu2023chatgpt} consider ChatGPT as a self-contained recommender system and construct a benchmark to track its performance in specific recommendation tasks, such as rating prediction and direct recommendation. ChatGPT can also serve as a component of existing recommender systems. Gao \emph{et al.} \cite{16gao2023chat} introduce Chat-REC, which employs ChatGPT as an interface for conversational recommendations, thereby enhancing the performance of existing recommendation models and rendering the recommendation process more interactive and explainable. Dai \emph{et al.}\cite{13dai2023chataug} propose ChatAug that utilizes ChatGPT to rephrase sentences for textual data augmentation, simultaneously demonstrating the effectiveness of ChatGPT as a text summarization tool when accompanied by pretrained language models (BERT). 

In terms of natural language generation tasks, ChatGPT demonstrates remarkable proficiency in generating persuasive recommendation interpretations and advertisements under specific conditions\cite{40remountakis2023using} \cite{14liu2023chatgpt}. Related research also suggests that the engagement of ChatGPT could be a innovative solution to address few-shot learning challenges\cite{41di2023retrieval}. However, recent research\cite{14liu2023chatgpt} reveals that when employed independently in few-shot scenarios as a recommender system, ChatGPT’s performance falls short compared to a series of classical recommendation models across diverse recommendation tasks, such as top-N direct recommendations. The aforementioned studies inspire us to explore the utilization of ChatGPT as an explicit feedback processing method indirectly participating in few-shot recommendation scenarios.

\section{Representations Generation}
\subsection{Task Formulation}
ChatGPT is designed to excel in user-oriented tasks, enabling us to adopt prompting paradigms\cite{42liu2023pre} to target specific tasks without the need for fine-tuning. Drawing partly from relevant studies\cite{14liu2023chatgpt}, our experiment initially utilizes the well-established ChatGPT model, \emph{gpt-3.5-turbo}, to generate textual user and item representations by providing ChatGPT with tailored prompts. Each prompt consists of three components: review injection, task description, and format indicator. The review injection is designed to provide ChatGPT with a sequence of reviews from the same subject (a specific user or item). The task description aims to elucidate the input materials and establish clear task requirements. The format indicator serves to standardize response formats and constrain content scope. Additionally, we set a limiter when generating prompts to prevent them from exceeding the maximum token limit in ChatGPT. 

Given that the API interface of ChatGPT necessitates its invocation in a conversational format, we assume the template  ${\tau}$, which denotes the procedure for employing ChatGPT to generate a textual representation of a specific subject by utilizing its review collections. Formally, this can be expressed as 
\begin{equation}
{\tau}=[X] suffix  [Y]\label{eq}
\end{equation}
where \emph{Y} represents a slot subsequently filled by ChatGPT’s response, \emph{suffix} (i.e., task description and format indicator) identifies certain text specifically designed to guide ChatGPT in accomplishing representation generation, and input \emph{X} is a sequence of reviews \emph{r} pertaining to the specific subject, formally:  
\begin{equation}
X = \{r_1, r_2, r_3, ..., r_n\}.
\end{equation}

\subsection{Generate Textual Representations by using ChatGPT}

\begin{figure*}[t!]
\centering
\includegraphics[width=\linewidth]{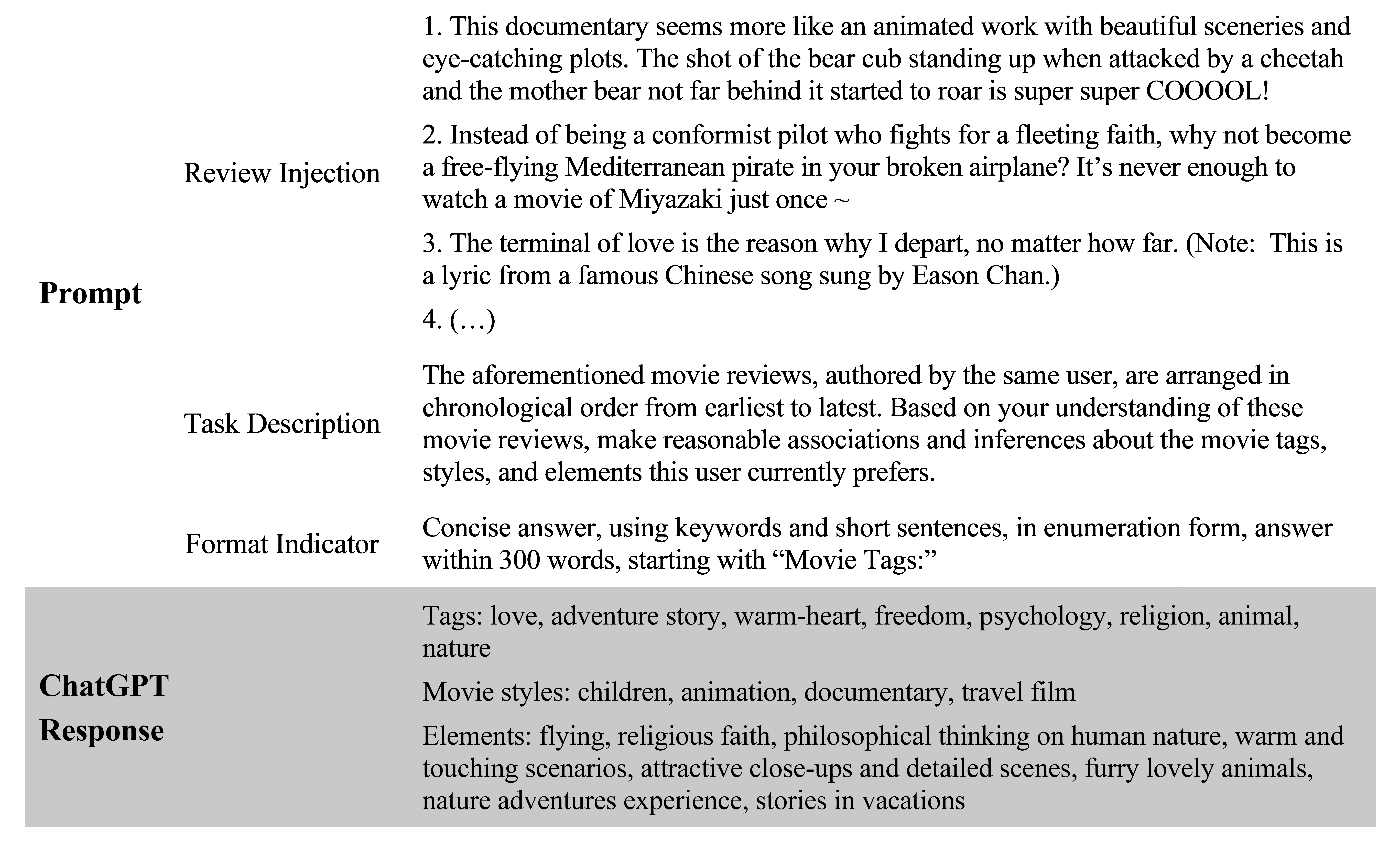}
\caption{Example of using ChatGPT to generate a textual user representation. Notably, the original reviews, prompts, and ChatGPT responses are all in Chinese; we employ ChatGPT to translate them into English for improved readability.}
\label{fig1}
\end{figure*}

The example in Fig.\ref{fig1} illustrates the generation of a user representation through template ${\tau}$. It is noteworthy that the generation of the item representations also adheres to template ${\tau}$, albeit with a slightly different \emph{suffix}; we modify the task description context for item representations generation to “(...Based on your understanding of these movie reviews, summarize the movie’s tag and scenes, associate and infer what type of audience and fans may be attracted by this movie.” In response to this description, ChatGPT would provide associations and inferences such as “Audiences who prefer heartwarming scenes and happy endings” for item representations. 

ChatGPT incorporates a certain degree of randomness to ensure the diversity of generated response, which may pose challenges in terms of reproduction and evaluation. The implementation of the format indicator component has been observed to effectively standardize the responses and mitigate irrelevant variations. During preliminary training with small sample sizes, ChatGPT exhibits exceptional association and inference capabilities that surpass our initial expectations. In certain instances, ChatGPT accurately “guesses” a specific movie and subsequently retrieves comprehensive information from its own database, even when the movie title is not explicitly mentioned in the original reviews. To ensure controlled variables, we explicitly instruct ChatGPT to exclusively focus on materials provided by us when generating representations. 

\subsection{Embed textual representations by using language models}
After generating textual user and item representations, we employ MacBERT\cite{43cui2020revisiting}, a pre-trained LLM for Chinese, to embed them to become our experimental dataset. Simultaneously, we construct a control dataset by concatenating reviews that belong to the same subject (item or user), embedding them with MacBERT, and merging the outputs. Additionally, we use a pre-trained Chinese Word2vec model\cite{44song2018directional} that does not employ attention mechanism to generate embeddings as an extra reference in some cases. 

The length of each embedding generated using MacBERT is 1,024, while the length of those generated using Word2vec is 200. Considering the superior efficiency of MacBERT in natural language embedding tasks, we primarily utilize MacBERT-processed embeddings as our main control datasets and only refer to experimental results obtained from using Word2vec-processed embeddings under specific conditions. The model selection as well as the embedding process partially drew upon a relevant study\cite{13dai2023chataug}.

\section{Evaluation}
To assess the effectiveness of LLMs as a textual explicit feedback processing method for recommender systems, we conduct ablation studies on diverse tasks with the aim of answer the following research questions: 
\begin{itemize}
\item RQ1: Do the LLM-processed user and item representations exhibit disparities compared to the original reviews?
\item RQ2: How effectively do these representation function across different recommendation models and tasks, in a few-shot scenario?
\item RQ3: Do the textual representations generated by ChatGPT in our experiment possess additional observable attributes and features, beyond those demonstrated in the aforementioned experiment results?
\end{itemize}

\subsection{Experimental Setup}
\subsubsection{Workflow}
\begin{figure*}[t!]
\centering
\includegraphics[width=\linewidth]{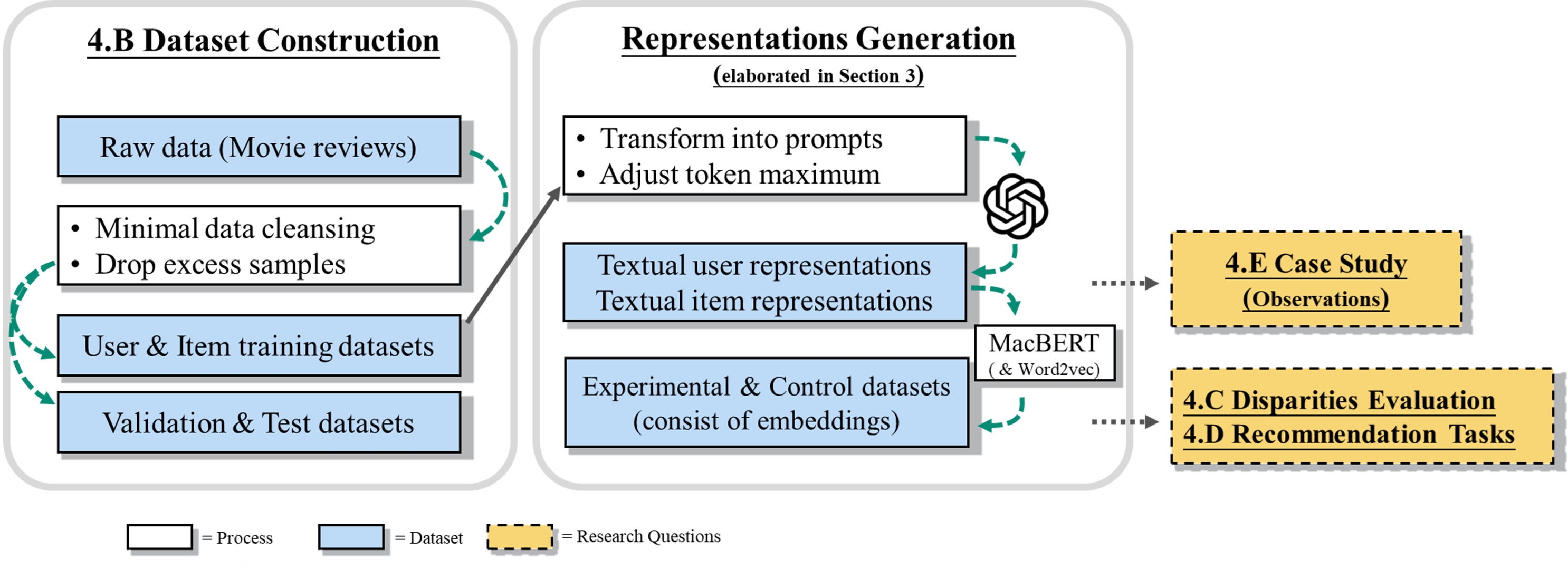}
\caption{Schematic representation of the complete  experimental workflow}
\label{fig2}
\end{figure*}

Building upon previous studies\cite{13dai2023chataug}\cite{14liu2023chatgpt}, we design our experimental workflow as follows: Firstly, we construct eligible datasets that include explicit user feedback and relevant information (Section 4.B). Secondly, the select user profiles and reviews are transformed into prompts for ChatGPT to generate textual user and item representations (elaborated in Section 3). Thirdly, the textual representations generated by ChatGPT undergo manual observation for case study purposes (Section 4.E) while concurrently being embedded by using MacBERT to construct an experimental dataset. Finally, the experimental dataset is incorporated into selected recommendation models for various recommendation tasks(Section 4.C, 4.D), along with control datasets. The complete workflow of our experimental process is illustrated in Fig.\ref{fig2}. 

\subsubsection{Baselines and Metrics}
In Section 4.C, we examine the disparities between the embeddings in the experimental dataset (ChatGPT-processed and MacBERT-embedded) and the embeddings in the control dataset (non-ChatGPT-processed and MacBERT-embedded). We employ three statistical methods\cite{45verma2020semantic}, namely cosine similarity, Manhattan distance, and Euclidean distance, to quantify the semantic relationships between embeddings of each subject (user/item) across the two datasets, namely \emph{embX} from the experimental datasets and \emph{$embX^{\prime}$} from the control datasets. We computed the mean cosine similarity, mean Manhattan distance, and mean Euclidean distance by averaging the results across all the subjects. The formula is presented below, where \emph{n} represents the size of the dataset and \emph{d} is the length of an individual embedding (1,024 for MacBERT embeddings):

\begin{equation}
mean\ Manhattan \ distance = \frac{1}{n}\sum_{k=1}^{n}\sum_{i=1}^{d}|embX_{ki}-embX^{\prime}_{ki}|
\end{equation}

\begin{equation}
mean\ Euclidean \ distance = \frac{1}{n}\sum_{k=1}^{n}\sqrt{\sum_{i=1}^{d}(embX_{ki}-embX^{\prime}_{ki})^{2}}
\end{equation}

\begin{equation}
mean \ cosine\ similarity  = \frac{1}{n}\sum_{k=1}^{n}\frac{\sum_{i=1}^{d}embX_{ki} embX ^{\prime}_{ki}}{\sqrt{\sum_{i=1}^{d} embX _{ki}^{2}} \sqrt{\sum_{i=1}^{d} embX _{ki}^{'2}}}.
\end{equation}

In Section 4.D, we evaluate the effectiveness of incorporating the LLM-processed embeddings into classical recommendation models for two recommendation tasks: interaction prediction and direct recommendation. The former constitutes a pivotal component in some neural network-based recommender systems\cite{46covington2016deep}\cite{47cheng2016wide}, while the latter represents a prevalent recommendation task. 

For interaction prediction (i.e., predicting whether a user will engage in interaction with a specific item), we employ Linear, MLP\cite{47cheng2016wide}, and CNN\cite{48kim2014convolutional}  models as our baselines. We consider user-item interactions as labels; specifically, ground truth interactions will be labeled as 1, while negative samples (labeled as 0) are generated by randomly assigning each user an item that they have not interacted with in reality. Given the binary classification nature of the task, we utilize Accuracy, Precision, and F1 Score as evaluation metrics to assess performance. For direct recommendation (i.e., recommending items that are most likely to align with a user's preferences), we employ BPR-MF\cite{49Frendle2012bpr}, NCF-Linear, NCF-MLP, and NCF-CNN\cite{50he2017neural} as baselines. 

We evaluate their performance using widely adopted metrics in recommender system studies, namely top-k Hit Ratio (HR@k) and top-k Mean Reciprocal Rank (MRR@k). Considering the few-shot scenario, we report results on either HR@{10,100} and either MRR@{10,100}. It is worth noting that despite varying in structural configurations, the aforementioned baselines integrating MLP and CNN neural networks have a comparable number of layers and are not fine-tuned respectively. 

\subsection{Dataset Construction}
The dataset employed in our experiment is the publicly available Douban Chinese Moviedata-10M\cite{51liu2021douban}, which shares similarities with the benchmark MovieLens dataset\cite{52harper2015movielens} in terms of content and format. The Douban dataset encompasses a substantial amount of explicit feedback provided by platform users, each sample presented as a user-item interaction comprising a user ID, an item ID, a piece of movie review, and other pertinent information such as a rating and a timestamp. In contrast to the MovieLens dataset, the Douban dataset primarily comprises Chinese text, and encompasses a substantial number of colloquial expressions, internet memes, emojis, and other intricate linguistic corpora. We intentionally perform minimal data cleansing to thoroughly evaluate ChatGPT’s proficiency in handling real-world explicit feedback. 

To construct our experimental dataset, a cohort of 1,000 users is randomly selected. We extract the historical user-item interaction samples of these users and sort them in chronological order. The item IDs corresponding to the two most recent interactions are extracted as test and validation samples, respectively, and are concatenated into their respective sets. The remaining interaction samples of these users constitute the training dataset for inputting into ChatGPT to generate textual user representations. To simulate a few-shot scenario, we artificially control the number of interaction samples per user by randomly discarding excess samples while ensuring at least one sample per user remains. Detailed statistical information about the user training dataset is provided in Tab.\ref{table1}.

\begin{table}[h]
\begin{threeparttable}
\caption{\textbf{User training dataset details}}
\label{table1}
\setlength{\tabcolsep}{3pt}

\begin{tabular}{p{50pt}p{145pt}p{30pt}}
\toprule
Dataset                                             & \multicolumn{2}{l}{Statistical   information}                 \\
\midrule
\multirow{2}{*}{User }                              & Interaction sample count                            & 7,270    \\
\multirow{2}{*}{representation}                     & Proportion   of users with samples \textless{}= 5   & 49.30\% \\
\multirow{2}{*}{training set}                       & Proportion of   reviews with words \textless{}= 100 & 81.03\% \\
                                                    & Proportion of   reviews with words \textless{}= 150 & 97.66\% \\
\bottomrule
\end{tabular}
\end{threeparttable}
\end{table}

After extracting all the samples corresponding to the aforementioned 1,000 users, we construct the remaining samples as the item training dataset. Each item in the dataset is designed to have at least one corresponding sample. We apply filtering rules to ensure that all items present in the user training dataset, validation dataset, and test dataset exist simultaneously in the item training dataset. Moreover, in order to maintain control variables when constructing the control datasets, we restrict the number of samples per item to a maximum of 10. Following filtration, approximately 98.05\% of the reviews in these samples are less than 150 words long (with 79.84\% being under 100 words).

\begin{figure*}[t!]
\centering
\includegraphics[width=0.8\linewidth]{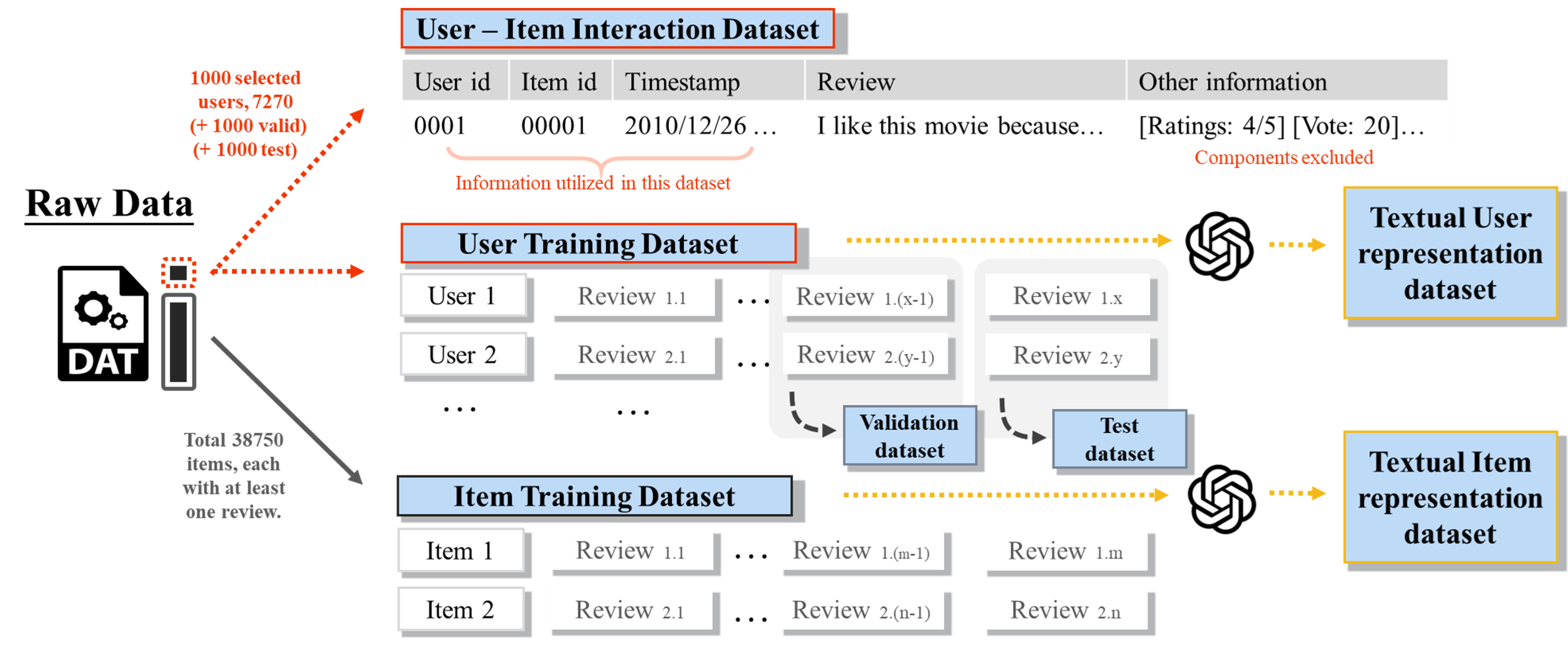}
\caption{Schematic representation of the datasets construction workflow}
\label{fig3}
\end{figure*}

Eventually, during the stage of constructing the training dataset, we obtain a user training dataset consisting of 1,000 users, encompassing a total of 7,270 interaction samples; additionally, we create validation and test datasets with each comprising 1,000 samples (one per user); finally, an item training dataset is compiled containing over 300,000 interaction samples from 38,750 items. By providing ChatGPT with tailored prompts derived from the two training datasets (elaborated in Section 3), we generate a corresponding number of textual user and item representations, which are then combined to form textual user and item representation datasets and subsequently embedded by language models to form experimental datasets. The workflow for constructing the datasets is illustrated in Fig.\ref{fig3}. 

As outlined in Section 3, we employ MacBERT and Word2vec to embed the textual item and user representations for generating embedding datasets. Additionally, we build control datasets in accordance with the methodology detailed in the section. 

In total, we acquire the following datasets: 
\begin{itemize}
\item A pair of textual representation datasets (user \& item).
\item A pair of experimental datasets (user \& item): ChatGPT-processed + MacBERT-embedded
\item Three pairs of control datasets (user \& item): Only MacBERT-embedded; Only Word2vec-embedded; ChatGPT-processed + Word2vec-embedded. 
\end{itemize}

\subsection{Disparities Evaluation (RQ1)}
In this section, we quantify the semantic relationships between embeddings of each subject (user/item) across the experimental dataset (ChatGPT-processed + MacBERT-embedded) and the control dataset (MacBERT-embedded). The evaluation method proposed in Section 4.A is employed to obtain the statistical measurement results presented in Tab.\ref{table2}.

\begin{table}[h]
\begin{threeparttable}
\caption{\textbf{Semantic distances between experimental and control datasets}}
\label{table2}
\setlength{\tabcolsep}{3pt}
\begin{tabular}{p{50pt}p{145pt}p{30pt}} 
\toprule
Dataset                                                                     & Statistical Methods     & Result \\
\midrule
\multirow{3}{*}{\begin{tabular}[c]{@{}l@{}}User \\ Embeddings\end{tabular}} & Mean Cosine similarity  & 0.94   \\
                                                                            & Mean Euclidean distance & 7.62   \\
                                                                            & Mean Manhattan distance & 194.09 \\
\multirow{3}{*}{\begin{tabular}[c]{@{}l@{}}Item \\ Embeddings\end{tabular}} & Mean Cosine distance    & 0.95   \\
                                                                            & Mean Euclidean distance & 6.84   \\
                                                                            & Mean Manhattan distance & 174.87 \\
\bottomrule                                                                            
\end{tabular}
\end{threeparttable}
\end{table}

The cosine similarity metric primarily focuses on the angular relationship between two vectors in a multi-dimensional space. When comparing two semantically similar sentences, regardless of their length, the angle between their vectors becomes smaller, resulting in a higher value for cosine similarity. Euclidean distance and Manhattan distance calculations encompass both direction and magnitude, which can serve as a complementary measure to cosine similarity.

When comparing the experimental and control datasets, both in terms of items and users, we observe that the result of Mean Cosine distance approaches 1, indicating a significant semantic similarity between the representations generated by ChatGPT and the original reviews. We also note that the Mean Euclidean and Manhattan distances deviate significantly from zero. Based on these results, we suggest that while the ChatGPT-generated representations demonstrate comparable semantics to the original reviews, they do exhibit significant disparities in terms of information content and quantity. This discrepancy may be attributed to their truncated length and refined content. In general, the aforementioned findings partially substantiate the effectiveness of ChatGPT in extracting a substantial portion of salient features and crucial information from the original reviews, albeit with a reconfigured textual composition and altered content. The reconfiguration and alteration will be examined in Section 4.E through a detailed case study.

\subsection{Performance Comparison on Recommendation Tasks (RQ2)}

\begin{figure*}[t!]
\centering
\includegraphics[width=0.8\linewidth]{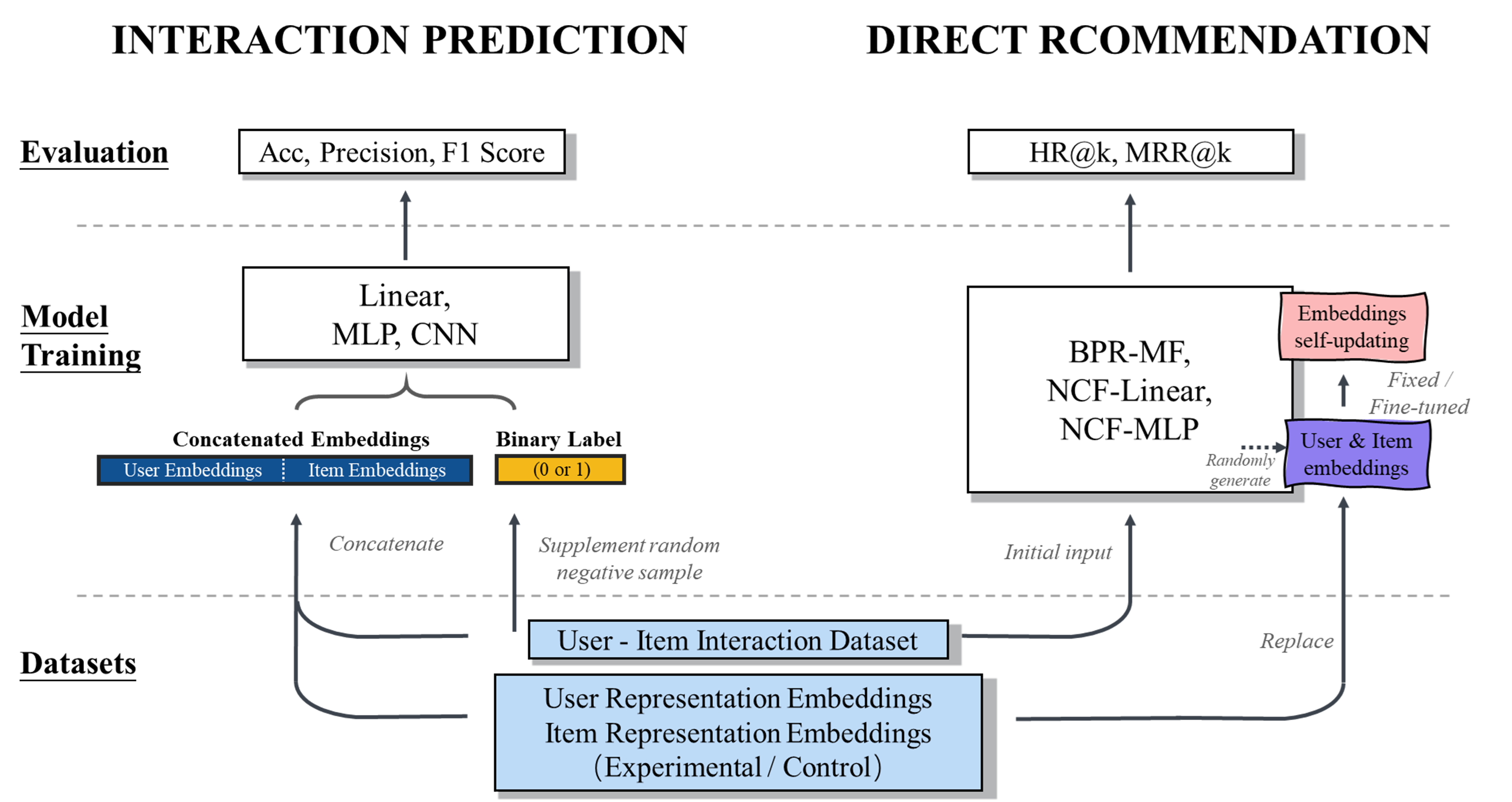}
\caption{Schematic representation of the experimental workflow for two recommendation tasks}
\label{fig4}
\end{figure*}

Fig.\ref{fig4} depicts the workflow for conducting ablation experiments on two recommendation tasks using user-item interactions and the user and item embeddings from both experimental and control datasets. Notably, we conduct 10 independent repetitions to train each model in the two recommendation tasks and report the average results, aiming to comprehensively investigate their overall performance.

For the interaction prediction task, we concatenate the user embedding and item embedding from the same interaction samples (including randomly generated negative samples) and input them into binary classification models along with their labels for training. Subsequently, we assess the model’s Accuracy, Precision, and F1 Score on the ground truth test dataset. 

For the direct recommendation task, we initialize the recommendation model with user-item interaction dataset. Upon model initialization, the BPR and NCF models automatically generate random embeddings for users and items, which are subsequently updated during training based on the models' learning from the user-item interaction dataset (with ratings). After model training, these fine-tuned embeddings serve as a foundation for recommending items to selected users. In our study, we eliminate ratings by substituting them with a uniform constant to prevent the recommendation model from relying on ratings. As compensation, we replace the model-automatically-generated embeddings with the user and item embeddings in our experimental and control datasets. We assess the performance of the models using HR (Hit Rate) and MRR (Mean Reciprocal Rank), while additionally considering scenarios where these embeddings continue to undergo fine-tuning or remain fixed during model training.

\subsubsection{Interaction prediction}

For the interaction prediction task, we conduct ablation experiments on experimental and control datasets using classical Linear, MLP, and CNN models respectively. The statistical measurements obtained from these experiments are reported in Tab.\ref{table3}. 

\begin{table}[h]
\begin{threeparttable}
\caption{\textbf{Performance comparison on interaction prediction tasks}}
\label{table3}
\setlength{\tabcolsep}{3pt}
\begin{tabular}{p{33pt}p{80pt}p{34pt}p{34pt}p{34pt}} 
\toprule
                         &                           & \multicolumn{3}{c}{Statistical Measurements}                            \\ \cmidrule(lr){3-5} 
\multirow{-2}{*}{Method} & \multirow{-2}{*}{Dataset} & Accuracy              & Precision             & F1 Score             \\
\midrule
                         & \textbf{ChatGPT + MacBERT} & \textbf{0.592}       & \textbf{0.601}        &\textbf{0.632}         \\
                         & Only MacBERT              & 0.552                 & 0.570                 & 0.523                 \\
                         & ChatGPT + Word2vec        & 0.501                 & 0.500                 & 0.500                  \\
\multirow{-4}{*}{MLP}    & Only Word2vec             & 0.500                 & 0.500                 & 0.500                 \\
Linear*                   & All datasets              & \multicolumn{3}{l}{\cellcolor[HTML]{F2F2F2}}                         \\
CNN*                      & All datasets              & \multirow{-2}{*}{\cellcolor[HTML]{F2F2F2} 0.500} & \multirow{-2}{*}{\cellcolor[HTML]{F2F2F2}0.250} & \multirow{-2}{*}{\cellcolor[HTML]{F2F2F2}0.335} \\
\bottomrule                                                                            
\end{tabular}
\begin{tablenotes}
\item[*] The Linear and CNN models exhibited unsuccessful convergence, with their Precision rate oscillating between either 0 or 0.5 and while F1 Score oscillating between either 0 or 0.67. We calculated the average of all experimental results and hereby provide an explanation.
\end{tablenotes}
\end{threeparttable}
\end{table}

Based on our observations, under the same MLP model, the experimental dataset demonstrate superiority over the control datasets. The results suggest that the incorporating ChatGPT-processed representation embeddings holds the potential to enhance certain recommender models that employ neural networks in a few-shot scenario. 

Notably, among all experimental models that integrated neural networks, the MLP model stands out as the only one to exhibit statistically significant results in both experimental and control datasets. In contrast, we observe that the CNN model exhibited a significantly high training loss and failed to successfully converge during training. We speculate that this phenomenon can be attributed to the length of the concatenated embedding and the limited number of the training samples, as certain neural networks may encounter detrimental effects on learning and convergence with a few-shot scenario characterized by an abundance of training features. This partially elucidates the unsatisfactory model performance observed in our experimental findings.

\subsubsection{Direct recommendation}

For the direct recommendation task, we conduct ablation experiments using experimental and control datasets on the BPR and NCF recommendation models, and investigate the impact of enabling or disabling automatic model updating during training. The specific experimental results are presented in Tab.\ref{table4} and Tab.\ref{table5}, with all outputs appropriately rounded to ensure a reader-friendly presentation.. Due to significant variations in performance among different recommendation models, we adopt HR and MRR @10 for NCF models and @100 for BPR models, respectively, to effectively showcase their performance. Furthermore, we present the percentage improvement of experimental models in comparison to the baseline model (which employs randomly generated embeddings) across diverse datasets, with a primary focus on results demonstrating an increase of 200\% or more for emphasis.

\begin{table}[h]
\begin{threeparttable}
\caption{\textbf{Performance comparison on BPR-MF model}}
\label{table4}
\setlength{\tabcolsep}{3pt}
\begin{tabular}{p{20pt}p{25pt}p{80pt}p{20pt}p{20pt}p{20pt}p{20pt}} 
\toprule

\multicolumn{2}{l}{\multirow{2}{*}{Method}}                                                                                                 & \multirow{2}{*}{Dataset} & \multicolumn{4}{c}{Statistical Measurements}              \\ \cmidrule(lr){4-7}
\multicolumn{2}{l}{}                                                                                                                        &                          & \multicolumn{2}{c}{HR@100} & \multicolumn{2}{l}{MRR@100} \\
\midrule
\multirow{9}{*}{\begin{tabular}[c]{@{}l@{}}BPR-\\ MF\end{tabular}} & \multirow{5}{*}{\begin{tabular}[c]{@{}l@{}}Fine-\\ tuned\end{tabular}} & ChatGPT + MacBERT        & 0.003         &              & 0.003         &               \\
                        &                                                                        & Only MacBERT             & 0.003         &              & 0.004                  & 200\%         \\
                        &                                                                        & \textbf{ChatGPT + Word2vec}       & \textbf{0.011}         & \textbf{550\%}        & \textbf{0.008}                  & \textbf{400\%}         \\
                        &                                                                        & Only Word2vec            & 0.008         & 400\%        & 0.006                  & 300\%         \\ \cmidrule(lr){3-7}
                        &                                                                        & ChatGPT + MacBERT        & 0.006         & 300\%        & 0.001         &               \\
                        & \multirow{3}{*}{Fixed}                                                 & Only MacBERT             & 0.006         & 300\%        & 0.001         &               \\
                        &                                                                        & ChatGPT + Word2vec       & 0.005         & 250\%        & 0.001         &               \\
                        &                                                                        & Only Word2vec            & 0.003         &              & 0.001         &               \\ \cmidrule(lr){3-7}
                        & Random                                                                 &                          & 0.002         & 100\%*       & 0.002         & 100\%*        \\
\bottomrule                                                                            
\end{tabular}
\begin{tablenotes}
\item[*] The table presents the significant results of the experimental models in comparison to the baseline model across diverse datasets, denoted as \%. 
\end{tablenotes}
\end{threeparttable}
\end{table}

\begin{table}[h]
\begin{threeparttable}
\caption{\textbf{Performance comparison on NCF models}}
\label{table5}
\setlength{\tabcolsep}{3pt}
\begin{tabular}{p{20pt}p{25pt}p{80pt}p{20pt}p{20pt}p{20pt}p{20pt}} %
\toprule
\multicolumn{2}{l}{\multirow{2}{*}{Method}}                                                                                                     & \multirow{2}{*}{Dataset} & \multicolumn{4}{c}{Statistical Measurements}              \\ \cmidrule(lr){4-7}
\multicolumn{2}{l}{}                                                                                                                            &                          & \multicolumn{2}{c}{HR@10} & \multicolumn{2}{l}{MRR@10} \\
\midrule
\multirow{5}{*}{\begin{tabular}[c]{@{}l@{}}NCF-\\ Linear\end{tabular}} & \multirow{2}{*}{\begin{tabular}[c]{@{}l@{}}Fine-\\ tuned\end{tabular}} & ChatGPT + MacBERT        & 0.041        &             & 0.003        &             \\
                                                                       &                                                                        & Only MacBERT             & 0.033        &             & 0.003        &             \\
                                                                       & \multirow{2}{*}{Fixed}                                                 & ChatGPT + MacBERT        & 0.080        & 267\%       & 0.004        & 200\%       \\
                                                                       &                                                                        & Only MacBERT             & 0.071        & 237\%       & 0.006        & 300\%       \\ 
                                                                       & Random                                                                 &                          & 0.030        & 100\%*      & 0.002        & 100\%*      \\ \cmidrule(lr){1-7}
\multirow{5}{*}{\begin{tabular}[c]{@{}l@{}}NCF-\\ MLP\end{tabular}}    & \multirow{2}{*}{\begin{tabular}[c]{@{}l@{}}Fine-\\ tuned\end{tabular}} & ChatGPT + MacBERT        & 0.092        &             & 0.006        &             \\
                                                                       &                                                                        & Only MacBERT             & 0.081        &             & 0.004        &             \\
                                                                       & \multirow{2}{*}{Fixed}                                                 & \textbf{ChatGPT + MacBERT}        & \textbf{0.210}        & \textbf{412\%}       & \textbf{0.012}        & \textbf{300\%}       \\
                                                                       &                                                                        & Only MacBERT             & 0.162        & 318\%       & 0.009        & 225\%       \\
                                                                       & Random                                                                 &                          & 0.051        & 100\%       & 0.004        & 100\%       \\ \cmidrule(lr){1-7}
\multirow{5}{*}{\begin{tabular}[c]{@{}l@{}}NCF-\\ CNN\end{tabular}}    & \multirow{2}{*}{\begin{tabular}[c]{@{}l@{}}Fine-\\ tuned\end{tabular}} & ChatGPT + MacBERT        & 0.080        &             & 0.006        &             \\
                                                                       &                                                                        & Only MacBERT             & 0.054        &             & 0.005        &             \\
                                                                       & \multirow{2}{*}{Fixed}                                                 & ChatGPT + MacBERT        & 0.104        & 248\%       & 0.013        & 260\%       \\
                                                                       &                                                                        & Only MacBERT             & 0.080        &             & 0.007        &             \\
                                                                       & Random                                                                 &                          & 0.042        & 100\%       & 0.005        & 100\%      \\
\bottomrule                                                                            
\end{tabular}
\begin{tablenotes}
\item[*] The table presents the significant results of the experimental models in comparison to the baseline models across diverse datasets and model structures, denoted as \%.
\end{tablenotes}
\end{threeparttable}
\end{table}

The ablation experiments demonstrate the significance of utilizing ChatGPT-processed embeddings to enhance a series of recommended models in few-shot scenarios. This enhancement is particularly evident in recommendation models that incorporate neural networks. Specifically, NCF-MLP outperforms NCF-CNN in terms of both HR and MRR metrics; models that fixed embeddings during training exhibit comparatively superior performance compared to those fine-tuned. 

Based on the experimental results, we suggest that the integration of neural networks enhances the recommendation models' capacity to process LLM-generated embeddings, which implies a substantial number of training features. 

We speculate that the limited sample size poses challenges for all neural networks, thereby compromising the validity of LLM-generated embeddings when automatically fine-tuned, whereas MLP is the sole network demonstrating superior adaptability in few-shot scenarios in our experiments (as evidenced by the results presented in the interaction prediction recommendation task). Meanwhile, recommendation models that do not incorporate neural networks encounter significant difficulties when dealing with lengthy embeddings. This could partially account for the superior experimental results obtained by utilizing Word2vec-embedded embeddings (which have shorter lengths compared to MacBERT-embedded embeddings) in BPR-MF models as opposed to other datasets.

\subsection{Case study (RQ3)}

In addition to conducting ablation experiments, we perform a comprehensive case study on the textual user and item representations to complement our findings and uncover potentially overlooked information within the embedding process. Our manual observations suggest that ChatGPT demonstrates exceptional proficiency in processing explicit textual feedback.

Specifically, it consistently demonstrates precise recognition and comprehension of contextual information with varying sentiment tendencies, even in the absence of quantitative metrics such as ratings. Notably, ChatGPT effectively handles reviews that contain positive, neutral, and negative snippets simultaneously by either disregarding the negative portion or considering an opposing viewpoint for recommendations. Additionally, ChatGPT adeptly identifies quotations within the reviews (e.g., movie lines, plots, extra materials) and utilizes them appropriately. The aforementioned observations collectively suggest that ChatGPT holds the potential to enhance the generalization capability of recommendation models by providing adaptability for diverse recommendation scenarios, such as social media platforms that exclusively comprising textual content.

Meanwhile, in contrast to conventional language models, ChatGPT demonstrates a unique ability to generate expansion context even when provided with limited information. While traditional NLP approaches primarily focus on keyword identification and extraction, ChatGPT goes beyond by introducing new content that may deviate from the original corpus. For instance, as depicted in Fig.\ref{fig1}, ChatGPT suggests the keyword "furry lovely animals," possibly due to the user's preference for documentaries featuring bears and animations. Essentially, ChatGPT “refines and reinforces” initial representations by augmenting them with supplementary information through association and inference. This could partially account for the observed semantic similarity yet content disparity between the experimental and control datasets, as evidenced by the findings in Section 4.C. Furthermore, the effectiveness of the refined and reinforced representations is demonstrated with support from the experimental results presented in Section 4.D. This partially indicates that the additional information contained within these representations, generated through ChatGPT's association and inference, carries significant implications. In other words, these supplementary pieces of information reasonably reflect users' underlying thoughts to a certain extent. To summarize, ChatGPT demonstrates its effectiveness in handling few-shot recommendation scenarios compared to conventional language models, owing to its distinctive capabilities in associative thinking and logical reasoning. 

It is noteworthy that in this experiment, ChatGPT functions as a symbolic representation of emerging LLMs endowed with generative and logical reasoning capabilities. Considering the continuous advancements in technology, forthcoming LLMs equipped with enhanced proficiencies in association and inference may ultimately supplant ChatGPT within our experimental framework. Nevertheless, the insights derived from this investigation retain significant reference value for future studies.

\section{Conclusion}
In this study, we conduct ablation experiments to assess the effectiveness of harnessing LLMs to enhance few-shot recommender systems in various recommendation tasks. Despite the limitations imposed by model structures, the inclusion of LLM-processed representations significantly enhances the performance of specific neural network-based recommendation models in our experimental few-shot scenario. Based on the experimental results, we suggest that LLMs equipped with generative and logical reasoning capabilities can serve as an effective NLP method for recommender systems, proficiently handling textual explicit feedback through their distinctive capabilities and enhancing the generalization potential of recommendation models. Moving forward, we envision integrating additional recommendation models based on neural networks into our study. Furthermore, we are intrigued by the potential business applications (e.g., marketing analytics, advertisement generation) of the ChatGPT-generated textual user and item representations.

\bibliographystyle{IEEEtran}
\bibliography{reference}

\section*{Acknowledgment}
Zhoumeng Wang wishes to extend his appreciation to Prof. Jimbo, Prof. Howard, Mr. Li Zhi, and Mr. Lei for their support throughout his academic journey.

\begin{IEEEbiography}[{\includegraphics[width=1in,height=1.25in,clip,keepaspectratio]{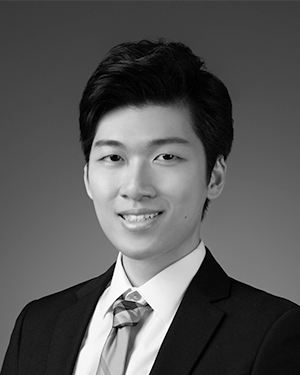}}]{Zhoumeng Wang}received his Master of Science degree in Marketing from The Chinese University of Hong Kong in 2023. He holds a bachelor's degree in Marketing from The Chinese University of Hong Kong, Shenzhen, which he obtained in 2022. His research interests lie in the fields of large language models, recommender systems, and digital marketing. Since his graduation, he has been involved as a research assistant.

He was awarded the prestigious Bowen Scholarship and consistently made the Dean's List twice during his study. In 2022, he participated in a Kaggle recommender system competition and earned a silver medal. 

\end{IEEEbiography}

\EOD

\end{document}